\documentclass[a4paper,12pt]{article}

\begin{document}
\begin{titlepage}
\title{Neutrino Oscillations mediated by the\\ Higgs field}
\author{W.\,Schmidt-Parzefall\\ \normalsize Universität Hamburg}
\end{titlepage}
\maketitle
\sloppy
\begin{abstract}Neutrino oscillations occur within the frame of the Standard Model, assuming that a neutrino is composed of a left handed and a right handed mass less fermion. Neutrino oscillations proceed via the 4-component Higgs field as intermediate state.\\

  Neutrino oscillations; Standard Model; Higgs field
\end{abstract}

\vspace{2cm}

Oscillations between quantum states may proceed via a mechanism involving an intermediate state. For example $B\bar{B}$ meson oscillations [1] proceed via a box graph dominated by the exchange of the $t$ quark. It is shown by the present paper that also neutrino oscillations [2] proceed via a particular graph occurring within the frame of the Standard Model after a minor modification of the description of neutrinos.\\

The Lagrangian $\mathcal{L}$ of the Standard Model for the observable particles [3] can be derived from a primary form of the Lagrangian [4], containing mass less fermions, mass less gauge bosons and the 4-component Higgs field [5]. A summary of this derivation is given by [6]. Neutrino oscillations are implied by the primary form of the Lagrangian. It consists of the following terms
\begin{equation}
\mathcal{L} = L_\psi + L_\chi + L_\phi + L_Y + L_W + L_B.
\end{equation}  
The terms of the primary form of $\mathcal{L}$ are
\begin{eqnarray}
L_\psi & = & \sum_i \overline{\psi_{iR}} \, \mbox{i} \gamma^\mu(\partial_\mu - U_\psi^{-1} \partial_\mu U_\psi) \psi_{iR} \nonumber \\
L_\chi & = & \sum_i \overline{\chi_{iL}} \,  \mbox{i} \gamma^\mu(\partial_\mu - U_\chi^{-1} \partial_\mu U_\chi) \chi_{iL} \nonumber \\
L_\phi & = & \frac{1}{2} \left( \left( (\partial_\mu - U_\phi^{-1} \partial_\mu U_\phi ) \phi \right)^\dagger ( \partial_\mu - U_\phi^{-1} \partial_\mu U_\phi ) \phi + \frac{m_H^2}{2} \phi^\dagger \phi - \frac{m_H^2}{4 v^2} (\phi^\dagger \phi)^2 \right) \nonumber \\
L_Y & = & - \sum_i C_i \left( \overline{\psi_{iR}} \, \phi^\dagger \chi_{iL} + \overline{\chi_{iL}}\, \phi \, \psi_{iR} \right) \nonumber \\
L_W & = & - \frac{1}{4} \bf{W}_{\mu \nu} \bf{W}^{\mu \nu} \nonumber \\
L_B & = & - \frac{1}{4} F_{\mu \nu} F^{\mu \nu} .
\end{eqnarray}

$\psi_{iR}$ is a mass less right handed fermion turning into the right handed component of an observable fermion with mass. It carries the weak isospin $t_3 = 0$ and the $U(1)$ charge $g\prime y_\psi$. It has $U(1)$ symmetry and interacts with the $U(1)$ potential $B_\mu$ by the interaction transformation
\begin{equation}
U_\psi = \exp \left( - \mbox{i} \int g\prime y_\psi B_\mu \mbox{d} x^\mu \right).
\end{equation}

$\chi_{iL}$ is a mass less left handed fermion turning into the lefthanded component of an observable fermion with mass. It carries the weak isospin $t_3 = \pm \frac{1}{2}$ and the $U(1)$ charge $g\prime y_\chi$. It has $SU(2) \times U(1)$ symmetry and interacts with the three $W_\mu$ potentials and the $B_\mu$ potential by the interaction transformation
\begin{equation}
U_\chi = \exp \left( - \mbox{i} \int \left( \frac{g}{\sqrt{2}} (T_+ W_\mu^+ + T_- W_\mu^-) + g t_3 W_\mu^3 + g\prime y_\chi B_\mu \right) \mbox{d} x^\mu \right),
\end{equation}  
where the $T_\pm$ are the weak isospin rising or lowering operators, and $g$ is the universal $SO(3)$ charge.\\

$\phi$ is a 4-component scalar field. $m_H \approx 125\,$GeV  is the mass of the observable 1-component Higgs particle, but the sign of the mass term of $\phi$ is opposite to the sign describing an observable particle. $v \approx 246\,$GeV is the electroweak scale.\\

The two mass less fermions $\psi_{iR}$ and $\chi_{iL}$ forming together a fermion with mass interact with each other via the field $\phi$ by a Yukawa interaction represented by the Lagrangian $L_Y$. The $C_i$ are coupling constants.\\

$L_Y$ describes the reactions
\begin{equation}
\psi_{iR} + \phi \rightarrow \chi_{iL} \quad \mbox{and} \quad \chi_{iL} \rightarrow \psi_{iR} + \phi .
\end{equation}
Since weak isospin and the $U(1)$ charge are conserved by these reactions, the interaction transformation of $\phi$ is
\begin{equation}
U_\phi = U_\psi^{-1} U_\chi . 
\end{equation}
The mass less fermions $\chi_{iL}$ and $\psi_{iR}$ carry electric charge. The electric charge $q$ is a linear combination of the weak charge $g t_3$ and the $U(1)$ charge $g\prime y$, realised by a rotation about the electroweak angle $\theta_W \approx 28.6^\circ$, which is defined by the condition, that the two mass less fermions forming together a fermion with mass, carry the same electric charge,\begin{equation}
q_\chi = \sin \theta_W g\,t_3 + \cos \theta_W g\prime y_\chi = \cos \theta_W g\prime y_\psi = q_\psi.
\end{equation} 
The electric charge of $\phi$ is
\begin{equation}
q_\phi = q_\chi -q_\psi = 0 . 
\end{equation}

The charges of the mass less fermions forming leptons are plotted in Figure 1, including the corresponding antiparticle states.\\

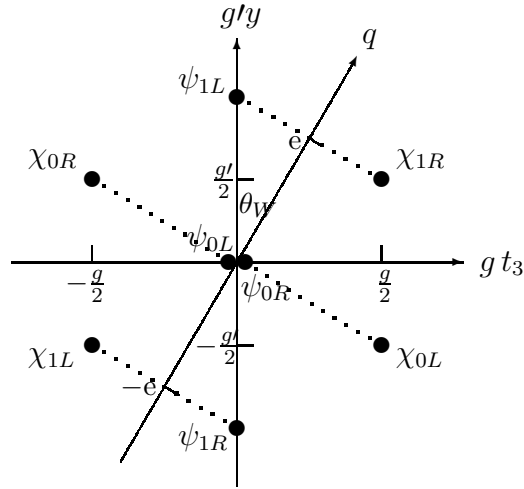
\begin{figure}[h]
\setlength{\unitlength}{1.1cm}
\begin{picture}(6,6)(-6.3,-2.7)
\put(-2.7,0){\vector(1,0){5.4}} \put(2.9,-0.1){$g\,t_3$}
\put(0,-2.7){\vector(0,1){5.4}} \put(-0.2,2.9){$g\prime y$}
\qbezier(-1.39,-2.4)(0,0)(1.39,2.4)
\put(1.34,2.3){\vector(1,2){0.1}} \put(1.5,2.65){$q$}
\small                        \put(0.02,0.6){$\theta_W$}
\put(-1.73,0){\line(0,1){0.2}}\put(-2.05,-0.35){$-\frac{g}{2}$}
\put(1.73,0){\line(0,1){0.2}} \put(1.66,-0.35){$\frac{g}{2}$}
\put(0,-1){\line(1,0){0.2}}   \put(-0.3,0.9){$\frac{g\prime}{2}$}
\put(0,1){\line(1,0){0.2}}    \put(-0.5,-1.1){$-\frac{g\prime}{2}$}
\qbezier(0.86,1.5)(0.94,1.45)(1.02,1.4) \put(0.62,1.37){e}
\qbezier(-0.86,-1.5)(-0.78,-1.55)(-0.7,-1.6) \put(-1.39,-1.6){$-$e}
\normalsize
\put(-1.73,-1){\circle*{0.2}} \put(-2.5,-1.2){$\chi_{1L}$}
\put(-1.73, 1){\circle*{0.2}} \put(-2.5, 1.2){$\chi_{0R}$}
\put(1.73, 1){\circle*{0.2}}  \put( 1.9, 1.2){$\chi_{1R}$}
\put(1.73,-1){\circle*{0.2}}  \put( 1.9,-1.2){$\chi_{0L}$}
\put(0,2){\circle*{0.2}}      \put(-0.7, 2.1){$\psi_{1L}$}
\put(0,-2){\circle*{0.2}}     \put(-0.7,-2.2){$\psi_{1R}$}
\put(-0.1,0){\circle*{0.2}}   \put(-0.6, 0.2){$\psi_{0L}$}
\put(0.1,0){\circle*{0.2}}    \put( 0.05,-0.4){$\psi_{0R}$}
\linethickness{0.04cm}
\qbezier[12](-1.73,-1)(-0.87,-1.5)(0,-2)
\qbezier[12](1.73,1)(0.87,1.5)(0,2)
\qbezier[12](-1.73,1)(-0.87,0.5)(-0.1,0)
\qbezier[12](1.73,-1)(0.87,-0.5)(0.1,0)
\end{picture}  
\it \caption{\label{Figure 1}The charges of the mass less fermions forming leptons approximate an SU(3) octet. The two mass less fermions $\chi_{iL}$ and $\psi_{iR}$ forming together a lepton are connected by a dotted line.} \rm
\end{figure}  

Deviating from the historical Standard Model, here a neutrino is assumed to be composed of a left handed and a right handed mass less fermion, like a charged lepton. The two constituents of a neutrino $\chi_L$ and $\psi_R$ interact with the Higgs field $\phi$ by a Yukawa interaction and form a massive fermion, obtaining its mass $m_i = C_i v$ by a term  of the Lagrangian $L_Y$.\\

The interactions implied by the Lagrangian $L_Y$ are described by the Feynman diagrams Figure 2.\\

\begin{figure}[h]
\setlength{\unitlength}{0.9cm}
\begin{picture}(14,5.5)(-7.8,-3)
\put(-5,2){\line(2,-1){2}} \put(-5,2){\vector(2,-1){1}} \put(-4.5,2){$\chi_L$}
\put(-5,-2){\line(2,1){2}} \put(-5,-2){\vector(2,1){1}} \put(-4.5,-2){$\psi_R$}
\put(-3,1){\line(2,1){2}}  \put(-3,1){\vector(2,1){1}}  \put(-2,2){$\psi_R$}
\put(-3,-1){\line(2,-1){2}}\put(-3,-1){\vector(2,-1){1}}\put(-2,-2){$\chi_L$}
\put(1,2){\line(1,-2){1}}  \put(1,2){\vector(1,-2){0.5}}\put(1.2,2){$\psi_R$}
\put(1,-2){\line(1,2){1}}  \put(1,-2){\vector(1,2){0.5}}\put(1.2,-2){$\chi_L$}
\put(4,0){\line(1,2){1}}   \put(4,0){\vector(1,2){0.5}} \put(4.3,2){$\chi_L$}
\put(4,0){\line(1,-2){1}}  \put(4,0){\vector(1,-2){0.5}}\put(4.2,-2){$\psi_R$}
\linethickness{0.1cm}      
\put(-3,-1){\line(0,1){2}} \put(-2.8,-0.1){$\phi$}
\put(2,0){\line(1,0){2}} \put(2.9,0.2){$\phi$}
\Large
\put(-5,0){$\nu_j$}\put(-1.5,0){$\nu_j$}\put(0.9,0){$\nu_j$}\put(4.6,0){$\nu_k$}
\large
\put(-3.3,-2.8){$2a$} \put(2.7,-2.8){$2b$}
\end{picture}  
\it \caption{\label{Figure 2}The Feynman diagrams corresponding to the Lagrangian $L_Y$ for the Yukawa interaction of the constituents $\chi_L$ and $\psi_R$ of a neutrino. The arrows denote ingoing and outgoing states. The diagram 2a is explicitly implied by $L_Y$. The diagram 2b occurs due to crossing symmetry. It transforms a $\nu_J$ into a $\nu_k$, $j,k = e,\mu,\tau$ and thus represents the mechanism for neutrino oscillations.} \rm
\end{figure}
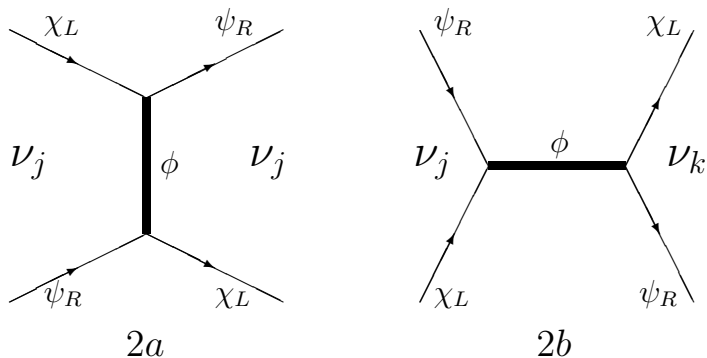  

Actually, only the reaction corresponding to the diagram $2a$ is explicitly implied by the Lagrangian $L_Y$. In order to describe the observed neutrino oscillations within the frame of the Standard Model, the assumption is made, that there is crossing symmetry [7], and therefore also the reaction of the rotated diagram $2b$ occurs. As a justification the similarity with the electromagnetic interaction is considered, where the $e^+ e^-$ Bhabha scattering proceeds via two diagrams related by crossing symmetry.\\

Conservation of weak isospin and $U(1)$ charge requires for the reaction $2b$ that the interaction transformation of the field $\phi$ is
\begin{equation} U_\phi = U_\psi U_\chi .\end{equation}
In addition, the relation (6) $U_\phi = U_\psi^{-1} U_\chi$ is imposed by diagram $2a$. Both conditions together result in the requirement
\begin{equation} U_\psi = 1. \end{equation}
This is fulfilled for neutrinos, but not for charged leptons or quarks. The $\psi_R$ component of neutrinos does not carry any charge. Therefore, the constituents of neutrinos interact not only according to diagram $2a$ but also according to diagram $2b$.\\

By both reactions, $2a$ and $2b$ a neutrino is transformed into a neutrino. By reaction $2a$ a neutrino is only transformed into itself. However, by reaction $2b$ a neutrino can also be transformed into a neutrino of a different generation. This results in neutrino oscillations, proceeding via the 4-component Higgs field $\phi$ as an intermediate state.\\

In summary, the observed neutrino oscillations can be explained within the frame of the Standard Model, assuming that a neutrino is composed of a left handed and a right handed mass less fermion, and assuming crossing symmetry for the reactions implied by the Yukawa interaction of the mass less fermions with the 4-component Higgs field. This picture allows to compute the neutrino oscillation rates, once the neutrino masses are known.\\

\newpage

{\large \bf References} 

\begin{tabbing}
[1] \=H. Albrecht et al. (ARGUS Coll.) Phys. Lett. {\bf B 192}, 245 (1987).\\

[2] R. Davis, D. Harmer, K. Hoffman, Phys. Rev. Lett. {\bf 20}, 1205 (1968).\\

\> Y. Fukuda et al. (Super-Kamiokande Coll.) Phys. Rev. Lett. {\bf 81}, 1562 (1998).\\

\> Q. Ahmad et al. (SNO Coll.) Phys. Rev. Lett. {\bf 87}, 071301 (2001).\\

\> N. Agafonova et al. (OPERA Coll.) Phys. Lett. {\bf B 691}, 138 (2010).\\

\> Y. Abe et al. (Double Chooz Coll.) Phys. Rev. Lett. {\bf 108} 131801 (2012).\\

\> S. Kim et al. (RENO Coll.) Phys. Rev. Lett. {\bf 108} 191802 (2012).\\

\> K. Abe et al. (T2K Coll.) Phys. Rev. {\bf D 88} 032002 (2013).\\

[3] Particle Data Group, Phys. Lett. {\bf B 667} 125 (2008),\\

\> and references therein.\\

[4] C.N. Yang and R.L. Mills, Phys. Rev. {\bf 69} 191 (1954).\\

\> S.L. Glashow, Nucl. Phys. {\bf 22} 579 (1961).\\

\> S. Weinberg, Phys. Rev.Lett. {\bf 19} 1264 (1967).\\

\> A. Salam ed. N. Swartholm, (Almquist and Wiksell, Stockholm, 1968).\\

\> G.`t Hooft and M. Veltman, Nucl. Phys. {\bf B 44} 189 (1972).\\

[5] F. Englert, R. Brout, Phys. Rev. Lett. {\bf 13} 321 (1964).\\

\> P.W. Higgs, Phys. Lett. {\bf 12} 132 (1964).\\

\> P.W. Higgs, Phys. Rev. Lett. {\bf 13} 508 (1964).\\

\> P.W. Higgs, Phys. Rev. {\bf 145} 1156 (1964).\\

\> G. Guralnik, C. Hagen, T.W.B. Kibble, Phys. Rev. Lett. {\bf 13} 585 (1964).\\

\> T.W.B. Kibble, Phys. Rev. {\bf 155} 1554 (1967).\\

[6] W. Schmidt-Parzefall, arXiv: 1406.1412 (2014).\\

[7] M. Gell-Mann and M. L. Goldberger, Phys. Rev. {\bf 96} 1433 (1954).
\end{tabbing}
\end{document}